\documentstyle[preprint,aps,epsfig]{revtex}
\tighten
\begin{document}
\preprint{Preprint Numbers: \parbox[t]{45mm}{ANL-PHY-8846-TH-97\\
}}

\title{Heavy- to light-meson transition form factors}

\author{M. A. Ivanov\footnotemark[1], 
Yu. L. Kalinovsky\footnotemark[2], 
P. Maris\footnotemark[3] and
C. D. Roberts\footnotemark[3]\vspace*{0.2\baselineskip}} 
\address{\footnotemark[1]Bogoliubov Laboratory of Theoretical Physics, \\
Joint Institute for Nuclear Research, 141980 Dubna,
Russia\vspace*{0.2\baselineskip}\\ 
\footnotemark[2]Laboratory of Computing Techniques and Automation, \\
Joint Institute for Nuclear Research, 141980 Dubna,
Russia\vspace*{0.2\baselineskip}\\ 
\footnotemark[3]Physics Division, Bldg. 203, Argonne National Laboratory,
Argonne IL 60439-4843\vspace*{0.2\baselineskip} }
\date{11th November 1997}
\maketitle
\begin{abstract}
Semileptonic heavy $\to$ heavy and heavy $\to$ light meson transitions are
studied as a phenomenological application of a heavy-quark limit of
Dyson-Schwinger equations.  Employing two parameters: $E$, the difference
between the mass of the heavy meson and the effective-mass of the heavy
quark; and $\Lambda$, the width of the heavy-meson Bethe-Salpeter amplitude,
we calculate $f_+(t)$ for all decays on their entire kinematically accessible
$t$-domain.  Our study favours $f_B$ in the range $0.135$-$0.17\,$GeV and
with $E=0.44\,$GeV and $1/\Lambda = 0.14\,$fm we obtain $f_+^{B\pi}(0) =
0.46$.  As a result of neglecting $1/m_c$-corrections, we estimate that our
calculated values of $\rho^2 = 0.87$ and $f_+^{DK}(0)=0.62$ are too low by
approximately 15\%.  However, the bulk of these corrections should cancel in
our calculated values of $Br(D \to \pi \ell \nu)/Br(D\to K \ell \nu)=0.13$
and $f_+^{D\pi}(0)/f_+^{DK}(0) = 1.16$.
\end{abstract}

\section{Introduction}
\label{intro}
Semileptonic meson decays are simple, experimentally accessible and only have
a single hadron in the initial and final state.  They are flavour-changing
weak interaction processes and hence can be used as a means of extracting the
elements of the Cabibbo-Kobayashi-Maskawa (CKM) matrix, which characterise
the difference between the mass eigenstates and the weak eigenstates in the
Standard Model.  For example, the $K^+\to \pi^0 e^+ \nu_e$ $(K_{e3}^+)$ and
$K_L^0 \to \pi^\pm e^\mp \nu_e$ $(K_{e3}^0)$ decays currently provide the
most accurate determination of $|V_{us}| (= 0.2196 \pm 0.0023)$~\cite{pdg96},
which would have been $\sin\theta_c$ in the Cabibbo theory of weak
interactions.  The mechanism of the weak interaction is well understood.
Hence, like elastic, electromagnetic form factors, these decays can also be
used as a tool to probe the structure of the hadrons in the initial and final
state.

A major goal of current $B$-meson experiments is to determine accurately the
matrix elements $V_{cb}$ and $V_{ub}$, the first of which is accessible via $
B \to D(D^*) \ell \nu$ decays and the latter via $ B \to \pi (\rho) \ell
\nu$.  The decays with a pseudoscalar meson in both the initial and final
state are the simplest to study theoretically because they are only sensitive
to the vector coupling of the $W$-boson to the quarks and only two form
factors are needed for a complete description.  However, experimentally those
with a vector meson in the final state provide the best statistics because
the decay can proceed via both $S$- and $D$-waves.

The $ B \to D(D^*) \ell \nu$ decays proceed via a $b\to c$ transition and
experimentally this is the closest one can come to realising a ``heavy $\to$
heavy'' transition~\cite{richman}.  It is in the analysis of these decays
that heavy-quark symmetry~\cite{neubert94}; i.e., an expansion of observables
in $\Lambda_{\rm QCD}/m_f$, where $m_f$ is the current-quark mass of the
$f=b,c$ quark, is most likely to be of use.  However, in reality the
$\Lambda_{\rm QCD}/m_c$-corrections, in particular, may nevertheless be large
$(\sim 30)$\% and difficult to estimate in this case.

Heavy Quark Effective Theory (HQET)~\cite{neubert94} provides a systematic
method for exploring the consequences of heavy-quark symmetry.  It can be
used to reduce the number of independent form factors required to describe
heavy $\to$ heavy decays, relating them to a minimal number of so-called
``universal'' form factors.  However, it can't be used to calculate the
$q^2$-dependence of the form factors.  This depends on the internal structure
of the heavy mesons and its calculation requires the application of
nonperturbative techniques.  One such technique and its application to the
calculation of these form factors is our focus herein.

The methods of HQET are also not directly applicable to the decays $B\to
\pi(\rho) \ell \nu_\ell$, $D\to K \ell \nu_\ell$ and $D \to \pi \ell
\nu_\ell$, all of which have light mesons in the final state.  A primary
impediment is that the current-quark mass of the $s$-quark $m_s \sim
O(\Lambda_{\rm QCD})$, hence $\Lambda_{\rm QCD}/m_s$ is not a suitable
expansion parameter.  In addition, a theoretical description of these decays
requires a good understanding of light quark propagation characteristics and
the internal structure of light mesons.  This is provided by the extensive
body of Dyson-Schwinger equation (DSE) studies~\cite{dserev,pctrev} in QCD.

The DSEs are a system of coupled integral equations whose solutions, the
$n$-point Schwinger functions, are the fully-dressed Euclidean propagators
and vertices for the theory.  Once all the Schwinger functions are known then
the theory is completely specified.  To arrive at a tractable problem one
must truncate the system at a given level.  Truncations that preserve the
global symmetries of a field theory are easy to implement~\cite{brs96}.
Preserving the gauge symmetry is more difficult but progress is being
made~\cite{ayse97}.

In a general covariant gauge the dressed-gluon 2-point Schwinger function
(Euclidean propagator), $D_{\mu\nu}(k)$, is characterised by a single scalar
function, which we denote ${\cal G}(k^2)/k^2$.  Important here is the
particular, qualitatively robust result of studies of the DSE for
$D_{\mu\nu}(k)$ that ${\cal G}(k^2)/k^2$ is strongly enhanced in the
infrared; i.e, its behaviour in the vicinity of $k^2=0$ can be represented as
a distribution~\cite{bp89,mikep}.  The infrared enhancement in
$D_{\mu\nu}(k)$ becomes prominent for $k^2\sim 1\,$GeV$^2$ and is not
peculiar to covariant gauges~\cite{otherIR}.

The dressed-quark propagator for a quark of flavour $f$ can be written in the
general form\footnote{We employ a Euclidean metric with
$\delta_{\mu\nu}={\rm diag}(1,1,1,1)$, $\gamma_\mu^\dagger = \gamma_\mu$ and
$\{\gamma_\mu, \gamma_\nu\}=2 \delta_{\mu\nu}$.  A spacelike 4-vector,
$k_\mu$, has $k^2>0$.}
\begin{equation}
\label{SZM}
S_f(p)= \frac{Z_f(p^2)}{i\gamma\cdot p  + M_f(p^2)}\,,
\end{equation}
where $Z_f(p^2)$ is the momentum-dependent wavefunction renormalisation and
$M_f(p^2)$ is the momentum-dependent quark mass-function.  The dressed-gluon
propagator is an important element in the kernel of the DSE satisfied by
$S_f(p)$.  In existing studies of this DSE that employ a dressed-quark-gluon
vertex that is free of kinematic, light-cone singularities, the infrared
enhancement in $D_{\mu\nu}(k)$ is sufficient to ensure that $S(p)$ doesn't
have a Lehmann representation.  This entails the absence of coloured quark
states from the spectrum; i.e., quark confinement~\cite{confinement}.  If
${\cal G}(k^2)<\infty$ at $k^2=0$ it is possible to obtain a solution,
$S_f(p)$, of the quark DSE that has a Lehmann
representation~\cite{hawes94}.

There is another important consequence of the infrared enhancement in ${\cal
G}(k^2)/k^2$.  The enhancement is characterised by a mass-scale $\omega \sim
\Lambda_{\rm QCD}$ and for light quarks; i.e, $u$-, $d$- and $s$-quarks for
which $m_f\leq \Lambda_{\rm QCD}$, it generates a significant enhancement in
$M_f(p^2)$.  A single, indicative and quantitative measure of this
enhancement in $M_f(p^2)$ is the ratio $M_f^E/m_f$, where $M_f^E$ is the
Euclidean constituent-quark mass defined as the solution of
$p^2=M^2(p^2)$.\footnote{Quark confinement entails that there is no
``pole-mass''~\cite{confinement}, which would be the solution of
$p^2+M^2(p^2)=0$.}  The results
\begin{equation}
\frac{M_{u,d}^E}{m_{u,d}} \sim 150 \,, \;\;  \frac{M_s^E}{m_s} \sim 10
\end{equation}
demonstrate that the infrared enhancement in ${\cal G}(k^2)/k^2$ leads to at
least an order-of-magnitude infrared enhancement in $M_f(p^2)$.  It is
nonperturbative in origin\footnote{The renormalisation-point-dependence of
the current-quark mass affects the actual value of the ratio $M_f^E/m_f$ but
not the qualitative features of this discussion.} and has important
qualitative and quantitative implications for light meson observables, as
illustrated in Refs.~\cite{pctrev,pich97}.

The effect of the infrared enhancement in ${\cal G}(k^2)/k^2$ on
$M_{c,b}(p^2)$ is much less dramatic~\cite{pmhl}:
\begin{equation}
\frac{M_b^E}{m_b} \sim 1.5\,, \;\; \frac{M_c^E}{m_c} \sim 2.0\,.
\end{equation}
In this case $m_f \gg\Lambda_{\rm QCD}$ and the momentum-dependence of
$M_{c,b}(p^2)$ is primarily perturbative in origin.  As observed in
Ref.~\cite{mishaA} it is therefore a good approximation to write 
\begin{equation}
\label{mbconst}
M_b(p^2)=\,{\rm const} := \hat M_b\approx M_b^E\,,
\end{equation}
for $p^2 \gg -m_b^2$, although the $b$-quark is still confined and there is no
pole mass.  For the same reason $Z_b(p^2) \equiv 1$ is also a good
approximation.  This and Eq.~(\ref{mbconst}) form the basis of the
heavy-quark limit of the DSEs explored in Ref.~\cite{mishaA} wherein, on the
domain explored by heavy $\to$ heavy semileptonic decays, the
dressed-$b$-quark propagator was approximated by
\begin{equation}
\label{sb}
S_{b}(p) = \frac{1}{i \gamma\cdot p + \hat M_b}\,.
\end{equation}
In Ref.~\cite{mishaA} the dressed-$c$-quark propagator was approximated by an
analogous expression:
\begin{equation}
\label{sc}
S_{c}(p) = \frac{1}{i \gamma\cdot p + \hat M_c}\,.
\end{equation}
However, the justification of this is less certain because the momentum
dependence of $Z_c(p^2)$ and $M_c(p^2)$ is significantly more rapid.  The
approach employed in Ref.~\cite{cjb97} is one means of exploring the fidelity
of this approximation, as are the direct studies for which Ref.~\cite{pmhl}
is the pilot.

Our aim herein is a unified description and correlation of semileptonic heavy
$\to$ heavy and heavy $\to$ light meson transitions as an extension of the
application of DSE methods.  We follow Ref.~\cite{mishaA} in describing the
$b$- and $c$-quark propagators by Eqs.~(\ref{sb}) and (\ref{sc}),
respectively, and in our analysis we consider the effects and limitations of
Eq.~(\ref{sc}).  These equations represent the primary, exploratory
hypothesis in our study because the propagation characteristics of
light-quarks and the structure of light meson bound states is well understood
following the extensive application of DSE methods in this
domain~\cite{pctrev,pich97,mr97}.  In Sec.~II we define our approximation to
the matrix elements describing $B(D) \to \pi(K) \ell \nu$ transitions and
fully specify a heavy-quark limit of our DSE application.  Our results are
presented and discussed in Sec.~III and we make some concluding remarks in
Sec.~IV.

\section{Semileptonic Decays}
Our primary focus is the pseudoscalar $\to$ pseudoscalar semileptonic decay
\begin{equation}
P_{H_1}(p_1) \to P_{H_2}(p_2)\, \ell \,\nu\,,
\end{equation}
where $P_{H_1}$ represents either a $B$ or $D$ meson with momentum $p_1$
$(p_1^2= -m_{H_1}^2)$ and $P_{H_2}$ can be a $D$, $K$ or $\pi$ meson with
momentum $p_2$ $(p_2^2= -m_{H_2}^2)$.  The momentum transfer to the lepton
pair is $q := p_1 - p_2$.  A review of these decays is provided in
Ref.~\cite{richman} and a theoretical study of the light $\to$ light
transitions is presented in Ref.~\cite{kpi}.

The invariant amplitude describing the decay is
\begin{equation}
A(P_{H_1} \to P_{H_2}\ell\nu) = 
\frac{G_F}{\surd 2} \,V_{qQ} \,
\bar\ell \gamma_\mu (1 -\gamma_5)\nu\, M_\mu^{P_{H_1} P_{H_2}}(p_1,p_2)\,,
\end{equation}
where $G_F$ is the Fermi weak-decay constant, $V_{qQ}$ is the appropriate
element of the CKM matrix and the hadronic current is
\begin{eqnarray}
M_\mu^{P_{H_1} P_{H_2}}(p_1,p_2) & := &
\langle P_{H_2}(p_2)| \bar q \gamma_\mu Q | P_{H_1}(p_1)\rangle\\
\label{fpfm}
& = & f_+(t) (p_1 + p_2)_\mu + f_-(t) q_\mu\,,
\end{eqnarray}
with $t := - q^2$.  The form factors, $f_\pm(t)$, contain all the information
about strong interaction effects in these processes and their accurate
estimation is essential to the extraction of $V_{qQ}$ from a measurement of a
semileptonic decay rate:
\begin{equation}
\label{branching}
\Gamma(P_{H_1} \to P_{H_2}\ell\nu)= 
\frac{G_F^2}{192 \pi^3}\,|V_{qQ}|^2\,\frac{1}{m_{H_1}^3}\,
\int_0^{t_-}\,dt\,|f_+(t)|^2\,
\left[(t_+-t) (t_- - t)\right]^{3/2}\,,
\end{equation}
with $t_\pm := (m_{H_1}\pm m_{H_2})^2$ and neglecting the lepton mass.

\subsection{Impulse Approximation}
In impulse approximation
\begin{eqnarray}
\label{ia}
\lefteqn{ M_\mu^{P_{H_1} P_{H_2}}(p_1,p_2) = }\\
&& \nonumber
\frac{N_c}{16\pi^4}\,
\int d^4k\,
{\rm tr}\left[
\bar\Gamma_{H_2}(k;-p_2) 
S_q(k+p_2) 
i {\cal V}_\mu^{qQ}(k+p_2,k+p_1)
S_Q(k+p_1)
\Gamma_{H_1}(k;p_1) 
S_{q^\prime}(k)\right]\,,
\end{eqnarray}
where: $\Gamma_{H_1}(k;p_1)$ is the Bethe-Salpeter amplitude for the $H_1$
meson; 
\begin{equation}
\bar\Gamma_{H_2}(k;-p_2)^t:= C^\dagger \Gamma_{H_2}(-k;-p_2)C,\;
C=\gamma_2\gamma_4, 
\end{equation}
and $M^t$ is the matrix transpose of $M$; and ${\cal V}_\mu^{qQ}(k_1,k_2)$ is
the vector part of the dressed-quark-W-boson vertex.

\subsubsection{Quark Propagators}
The dressed quark propagators, $S_f(p)$, in Eq.~(\ref{ia}) are the solution
of 
\begin{eqnarray}
\label{gendse}
S(p)^{-1} & = & i\gamma\cdot p + m_{\rm bm}
+ \int\frac{d^4 q}{(2\pi)^4} \,
g^2 D_{\mu\nu}(p-q) \frac{\lambda^a}{2}\gamma_\mu S(q)
\Gamma^a_\nu(q,p)\,,
\end{eqnarray}
where: $D_{\mu\nu}(k)$ is the dressed-gluon propagator; $\Gamma^a_\nu(q,p)$
is the dressed-quark-gluon vertex; $m_{\rm bm}$ is the current-quark bare
mass; and one can write $S_f(p)$ in the general form
\begin{equation}
\label{genS}
S_f(p) = -i\gamma\cdot p\, \sigma_V^f(p^2) + \sigma_S^f(p^2)\,,
\end{equation}
which is completely equivalent to Eq.~(\ref{SZM}).  A thorough discussion of
the numerical solution of Eq.~(\ref{gendse}), including a discussion of
renormalisation, is given in Ref.~\cite{mr97}.

\paragraph{Light quarks.}
Herein, for the light $u$-, $d$- and $s$-quark propagators, we do not directly
employ a numerical solution of Eq.~(\ref{gendse}).  Instead we use the
algebraic parametrisations of these solutions developed in Ref.~\cite{brt96}
because they efficiently characterise the essential and robust elements of
the solution obtained in many studies~\cite{dserev} of the quark DSE:
\begin{eqnarray}
\label{SSM}
\bar\sigma^f_S(x)  & =  & 
        2 \bar m_f {\cal F}(2 (x + \bar m_f^2))
        + {\cal F}(b_1 x) {\cal F}(b_3 x) 
        \left( b^f_0 + b^f_2 {\cal F}(\epsilon x)\right)\,,\\
\label{SVM}
\bar\sigma^f_V(x) & = & \frac{2 (x+\bar m_f^2) -1 
                + e^{-2 (x+\bar m_f^2)}}{2 (x+\bar m_f^2)^2}\,,
\end{eqnarray}
where: $f=u,s$ (isospin symmetry is assumed), 
\begin{equation}
{\cal F}(y):= \frac{ 1-{\rm e}^{-y}}{y }\,;
\end{equation}
$x=p^2/(2 D)$; $\bar m_f$ = $m_f/\sqrt{2 D}$; and
\begin{eqnarray}
\bar\sigma_S^f(x) & := & \sqrt{2 D}\,\sigma_S^f(p^2)\,,\\
\bar\sigma_V^f(x) & := & 2 D\,\sigma_V^f(p^2)\,,
\end{eqnarray}
with $D$ a mass scale.  This algebraic form combines the effects of
confinement and dynamical chiral symmetry breaking with free-particle
(asymptotically-free) behaviour at large, spacelike $p^2$.  The parameters
$\bar m_f$, $b_{0\ldots 3}^f$ in Eqs.~(\protect\ref{SSM}) and
(\protect\ref{SVM}) take the values
\begin{equation}
\label{tableA} 
\begin{array}{cccccc}
        & \bar m_f& b_0^f & b_1^f & b_2^f & b_3^f \\\hline
 u:  & 0.00897 & 0.131 & 2.90 & 0.603 & 0.185 \\
 s:  & 0.224   & 0.105 & \underline{2.90} & 0.740 & \underline{0.185}
\end{array}\,,
\end{equation}
which were determined in a least-squares fit to a range of light-hadron
observables.  The values of $b_{1,3}^s$ are underlined to indicate that the
constraints $b_{1,3}^s=b_{1,3}^u$ were imposed in the fitting.  The scale
parameter $D=0.160\,$GeV$^2$.  

\paragraph{Heavy quarks.}
As described in Sec.~\ref{intro}, and exploited in Ref.~\cite{mishaA}, the
momentum-dependence of $Z_f(p^2)$ and $M_f(p^2)$ is much weaker for the
heavy-quarks than it is for the light-quarks.  This is illustrated for two
different but related DSE-models in Refs.~\cite{pmhl,mishaA} and justifies
Eq.~(\ref{sb}) for the $b$-quark and the cautious, exploratory use of
Eq.~(\ref{sc}) for the $c$-quark.

These equations provide the origin of heavy-quark symmetry in the DSE
framework.  Its elucidation is completed by introducing the heavy-meson
velocity, $v_\mu$, via
\begin{equation}
\label{vel}
p_{1\mu} := m_{H_1}\,v_\mu := (\hat M_{f_Q} + E)\, v_\mu\,,
\end{equation}
where $v^2=-1$ and $E>0$ is the difference between the heavy-meson mass and
the effective-mass of the heavy-quark, $\hat M_{f_Q}$.  Equations~(\ref{sb})
and (\ref{sc}) then yield
\begin{equation}
\label{hqf}
S_{f_Q}(k+p_1) = \case{1}{2}\,\frac{1 - i \gamma\cdot v}{k\cdot v - E}
+ {\rm O}\left(\frac{|k|}{\hat M_{f_Q}},
                \frac{E}{\hat M_{f_Q}}\right)\,.
\end{equation}
Exact heavy-quark symmetry arises from completely neglecting the $1/\hat
M_{f_Q}$ corrections in all applications.  The mass of the $b$-quark may
justify this as a quantitatively reliable approximation but in making the
same truncation for the $c$-quark one may expect quantitatively important
corrections.

\subsubsection{Bethe-Salpeter amplitude}
As discussed in Refs.~\cite{pmhl,mr97}, the meson Bethe-Salpeter amplitudes
in Eq.~(\ref{ia}) are the solution of the homogeneous Bethe-Salpeter
equation:
\begin{eqnarray}
\label{genbse}
\left[\Gamma_H(k;P)\right]_{tu} &= & 
\int\frac{d^4 q}{(2\pi)^4} \,
[\chi_H(q;P)]_{sr} \,K^{rs}_{tu}(q,k;P)\,,
\end{eqnarray}
where 
\begin{equation}
\label{chiform}
\chi_H(q;P) \doteq S_Q(q+P) \Gamma_H(q;P) S_{q^\prime}(q)\,,
\end{equation}
$S_f$ are the dressed-quark propagators, and $r$,\ldots,$u$ represent
colour-, Dirac- and flavour-matrix indices.  In Eq.~(\ref{genbse})
$K^{rs}_{tu}(q,k;P)$ is the fully-amputated quark-antiquark scattering
kernel.  $K^{rs}_{tu}(q,k;P)$ is a $4$-point Schwinger function obtained as
the sum of a countable infinity of skeleton diagrams.  It is
two-particle-irreducible, with respect to the quark-antiquark pair of lines,
and does not contain quark-antiquark to single gauge-boson annihilation
diagrams, such as would describe the leptonic decay of a pseudoscalar meson.
The numerical studies of Ref.~\cite{mr97} employed a ladder-like
approximation:
\begin{equation}
K^{rs}_{tu}(q,k;P) = - g^2 D_{\mu\nu}(k-q)\,
        \left(\gamma_\mu\frac{\lambda^a}{2}\right)_{tr}\,
        \left(\gamma_\nu\frac{\lambda^a}{2}\right)_{su}\,,
\end{equation}
which is consistent with the impulse approximation for $M_\mu^{P_{H_1}
P_{H_2}}(p_1,p_2)$ and is a quantitatively reliable truncation for light,
pseudoscalar mesons because of cancellations, order-by-order, between higher
order diagrams in the skeleton expansion for $K$~\cite{brs96}.
Ref.~\cite{pmhl} is a first step in exploring the application of the methods
of Ref.~\cite{mr97} to mesons containing at least one heavy quark.

\paragraph{Heavy meson Bethe-Salpeter amplitudes.}
Herein we do not use a numerical solution of Eq.~(\ref{genbse}) for the
heavy-meson Bethe-Salpeter amplitude because we judge that our present
studies are inadequate.  One limitation, for example, is that simple
ladder-like truncations do not yield the Dirac equation when the mass of one
of the fermions becomes infinite and that defect may also be manifest in our
study.  Postponing the detailed exploration of this and other questions we
employ instead an Ansatz motivated by the studies of Ref.~\cite{bsesep} and
used efficaciously in Ref.~\cite{mishaA}:
\begin{equation}
\label{hmbsa}
\Gamma_{H_{1f}}(k;p_1) = 
\gamma_5 \left(1 + \case{1}{2} i \gamma\cdot v\right)
\case{1}{{\cal N}_{H_{1f}}}\,\varphi(k^2)\,,
\end{equation}
where ${\cal N}_{H_{1f}}$ is the canonical Bethe-Salpeter normalisation
constant.  Using Eq.~(\ref{hqf}) 
\begin{equation}
\label{bsanormH}
{\cal N}_{H_{1f}}^2 = \frac{1}{m_{H_{1f}}}\,\frac{N_c}{32 \pi^2}
\int_0^\infty du \,
\varphi(z)^2\,
\left(\sigma_S^f(z) + \sqrt{u} \,\sigma_V^f(z)\right)
:= \frac{1}{m_{H_{1f}} \kappa_f^2} \,,
\end{equation}
where $z= u - 2 E \sqrt{u}$ and $f$ labels the light-quark flavour.

In a solution of the Bethe-Salpeter equation the form of $\varphi(k^2)$ is
completely determined.  However, here it characterises our Ansatz and as our
primary form we choose
\begin{equation}
\label{phia}
\varphi(k^2) = \exp\left(-k^2/\Lambda^2\right)\,,
\end{equation}
where $\Lambda$ is a free parameter.  In studies of heavy $\to$ heavy
transitions~\cite{mishaA} we found that, as long as $\varphi(k^2)$ is a
non-negative, non-increasing, convex up function of $k^2$, the results were
insensitive to its detailed form.  As we shall see below, through a
comparison of the results obtained using Eq.~(\ref{phia}) and those obtained
with
\begin{equation}
\label{phib}
\tilde\varphi(k^2)= \frac{\tilde\Lambda^2}{k^2 + \tilde\Lambda^2}\,,
\end{equation}
the same is true herein.  Qualitatively, a primary requirement for an
understanding of all the processes we consider is simply that the heavy meson
be represented by a function that describes it as a finite-size, composite
object: $1/\Lambda$ is a rough measure of that size.

\paragraph{Light meson Bethe-Salpeter amplitudes.}
Just as for the light-quark DSE, there have been numerous
studies~\cite{dserev,pctrev} of light mesons using Eq.~(\ref{genbse}) and a
thorough discussion of the numerical solution, including a discussion of
renormalisation, is presented in Ref.~\cite{mr97}.  The light, pseudoscalar
meson Bethe-Salpeter amplitude has the general form
\begin{eqnarray}
\label{genpibsa}
\Gamma_H(k;P) & = &  \gamma_5 \left[ i E_H(k;P) + 
\gamma\cdot P F_H(k;P) \rule{0mm}{5mm}\right. \\
\nonumber & & 
\left. \rule{0mm}{5mm}+ \gamma\cdot k \,k \cdot P\, G_H(k;P) 
+ \sigma_{\mu\nu}\,k_\mu P_\nu \,H_H(k;P) 
\right]\,.
\end{eqnarray}
Until recently it was assumed that in quantitative phenomenological
applications one could neglect all but $E_H(k;P)$ in describing the light,
pseudoscalar meson and this was the assumption of Ref.~\cite{brt96}.
However, a systematic study of the quark DSE and meson Bethe-Salpeter
equation~\cite{mr97} demonstrates that the other functions are both
qualitatively and quantitatively important.  A reanalysis of elastic form
factors using all amplitudes and refitting the parameters characterising the
quark propagators is therefore necessary.  It is underway but
incomplete~\cite{mrpion}.

Herein we use the parametrisation of the light meson Bethe-Salpeter amplitude
determined in Ref.~\cite{brt96} and the results~\cite{mr97} that it is a good
approximation to treat $E_H(k;P)= E_H(k^2)$ and $E_\pi(k^2) = E_K(k^2):={\cal
E}(k^2)$; i.e., we use
\begin{eqnarray}
\Gamma_{H=\pi,K}(k^2) & = & i \gamma_5\,{\cal E}(k^2) \,,\\
{\cal E}(k^2) & = & 
\frac{\surd 2}{f_H}\,
\frac{C_0 \,{\rm e}^{-k^2/[2 D]} + \left.\sigma_S(k^2)\right|_{m_f=0}}
        {\left.\sigma_V(k^2)\right|_{m_f=0}}\,,
\end{eqnarray}
where the parameter $C_0= 0.214\,$GeV was fixed in Ref.~\cite{brt96} and
therein yields the experimental value $f_\pi= 0.131$.  For the kaon $f_K=
0.196\,$GeV.  

In principle, neglecting the other amplitudes in Eq.~(\ref{genpibsa}) is
flawed.  However, the light quark propagators of
Eqs.~(\ref{SSM})-(\ref{tableA}) were also fixed under this assumption and it
is the {\it combination} of these parametrisations in Eq.~(\ref{chiform})
that appears in the calculation of hadronic observables and reproduces
available data.  Therefore, if practiced judiciously, neglecting the other
amplitudes can still provide quantitatively reliable results.  To illustrate
this we note that a preliminary reanalysis of the electromagnetic pion form
factor~\cite{mrpion}, using all the amplitudes in Eq.~(\ref{genpibsa}) and
refitting the $u$-quark propagator parameters in Eq.~(\ref{tableA}), yields
results that are qualitatively indistinguishable from those obtained in
Ref.~\cite{brt96} for $q^2\leq 20\,$GeV$^2$.  It is only for $q^2>
20\,$GeV$^2$ that the qualitative and quantitative importance of $F_\pi$ and
$G_\pi$ becomes manifest: these are the dominant amplitudes at large $q^2$
and ensure that $q^2 F_\pi(q^2) = {\rm const}$, up to $\ln q^2$ corrections.

\subsubsection{Quark-W-boson vertex}
${\cal V}_\mu^{qQ}(k_1,k_2)$ in Eq.~(\ref{ia}) satisfies a DSE that describes
both the strong and electroweak dressing of the vector part of the
quark-W-boson vertex.  Solving this equation is a problem that can be
addressed using the methods of Ref.~\cite{mr97}.  However, we postpone this
problem for the present and note instead that from this DSE one can derive a
Ward-Takahashi identity
\begin{equation}
(k_1 - k_2)_\mu\,i{\cal V}_\mu^{f_1f_2}(k_1,k_2) = S_{f_1}^{-1}(k_1) -
S_{f_2}^{-1}(k_2) - (m_{f_1} - m_{f_2})\,\Gamma_I^{f_1f_2}(k_1,k_2)\,,
\end{equation}
where $\Gamma_I^{f_1f_2}(k_1,k_2)$ is the scalar vertex, which in the absence
of interactions is simply the diagonal unit matrix in Dirac space.  This
identity can be used to constrain the form of ${\cal
V}_\mu^{f_1f_2}(k_1,k_2)$, as the QED analogue has been used to constrain the
dressed-quark-photon vertex~\cite{vertex}.

When $f_1$ and $f_2$ are both heavy quarks then the ability to neglect gluon
dressing, as manifest in Eq.~(\ref{sb}), entails
\begin{equation}
(m_{f_1} - m_{f_2})\,\Gamma_I^{f_1f_2}(k_1,k_2)
\approx
(\hat M_{f_1} - \hat M_{f_2})\,1_D\,.
\end{equation}
This justifies the approximation, used efficaciously in Ref.~\cite{mishaA},
\begin{equation}
\label{bare}
{\cal V}_\mu^{f_1f_2}(k_1,k_2) = \gamma_\mu
\end{equation}
thereby amplifying the simplifications accruing in the heavy-quark limit.  As
demonstrated in Ref.~\cite{kpi}, even in the case where both quarks are
light, improvements to Eq.~(\ref{bare}) only become quantitatively
significant ($\sim 10$\%) in the magnitude of $f_+(t)$ at the extreme
kinematic limit: $t=t_-$.  Hence we use Eq.~(\ref{bare}) in all calculations
described herein.

\subsection{Semileptonic decays in the heavy-quark limit}
\subsubsection{$B_f \to D_f$}
Using Eqs.~(\ref{genS}), (\ref{hqf}) and (\ref{hmbsa}), we find~\cite{mishaA}
from Eqs.~(\ref{fpfm}) and (\ref{ia}) that, at leading order in $1/m_H$ where
$m_H$ is the heavy-meson mass,
\begin{eqnarray}
f_\pm(t) & = & \case{1}{2}\, 
\frac{m_{D_f} \pm m_{B_f}}{ \sqrt{m_{D_f} m_{B_f}} }\,\xi_f(w) \,,\\
\label{xif}
\xi_f(w) & = & \kappa_f^2\,\frac{N_c}{32\pi^2}\,
\int_0^1 d\tau\,\frac{1}{W}\,
\int_0^\infty du \, \varphi(z_W)^2\,
        \left[\sigma_S^f(z_W) + \sqrt{\frac{u}{W}} \sigma_V^f(z_W)\right]\,,
\end{eqnarray}
with $W= 1 + 2 \tau (1-\tau) (w-1)$, $z_W= u - 2 E \sqrt{u/W}$
and\footnote{The minimum physical value of $w$ is $w_{\rm min}=1$, which
corresponds to maximum momentum transfer with the final state meson at rest;
the maximum value is $w_{\rm max} \simeq (m_{B_f}^2 + m_{D_f}^2)/(2 m_{B_f}
m_{D_f}) = 1.6$, which corresponds to maximum recoil of the final state meson
with the charged lepton at rest.}
\begin{equation}
w = \frac{m_{B_f}^2 + m_{D_f}^2 - t}{2 m_{B_f} m_{D_f}} = v_{B_f} \cdot
v_{D_f}\,. 
\end{equation}
The canonical normalisation of the Bethe-Salpeter amplitude,
Eq.~(\ref{bsanormH}), automatically ensures that 
\begin{equation}
\xi_f(w=1) = 1\,.
\end{equation}
Equation~(\ref{xif}) is an example of a general result that, in the
heavy-quark limit, the semileptonic $H_f \to H_f^\prime$ decays of heavy
mesons are described by a single, universal function: $\xi_f(w)$~\cite{IW90}.

\subsubsection{ Heavy $\to$ Light}
Using Eqs.~(\ref{genS}), (\ref{hqf}) and (\ref{hmbsa}), and following the
method outlined in the appendix, we find from Eqs.~(\ref{fpfm}) and
(\ref{ia})
\begin{equation}
\label{fphl}
f_+^{H_1 H_2}(t) = \kappa_{q^\prime}
         \frac{\surd 2}{f_{H_2}}\,\frac{N_c}{32\pi^2}\,
        F_{q^\prime}(t;E,m_{H_1},m_{H_2}) \,,
\end{equation}
where
\begin{equation}
F_{q^\prime}(t;E,m_{H_1},m_{H_2}) =
\frac{4}{\pi}\,\int_{-1}^{1}\,\frac{d\gamma}{\sqrt{1-\gamma^2}}\,
        \int_0^1\,d\nu\,
        \int_0^\infty u^2 du\,\varphi(z_1)\,
        {\cal E}(z_1)\,W_{q^\prime}(\gamma,\nu,u) \,,
\end{equation}
with 
\begin{eqnarray}
W_{q^\prime}(\gamma,\nu,u) & = & 
2 \tau^2 \left[ \sigma_S^u(z_1) \frac{d}{dz_2}\sigma_V^{q^\prime}(z_2) 
       - \sigma_V^u(z_1)\frac{d}{dz_2}\sigma_S^{q^\prime}(z_2) \right]\\
\nonumber
& + & \left(1 - \frac{u \,\nu}{m_{H_1}}\right)\,
        \sigma_S^u(z_1)\,\sigma_V^{q^\prime}(z_2)\\
\nonumber
& + & \frac{1}{m_{H_1}}
\left[ \rule{0mm}{1.3\baselineskip}
        \sigma_S^u(z_1)\,\sigma_S^{q^\prime}(z_2)
        + u\,\nu\,\sigma_V^u(z_1)\,\sigma_S^{q^\prime}(z_2) \right. \\
\nonumber
&& \left. + \left(z_1 + u\,\nu\,M_{H_1}\right)
                \sigma_V^u(z_1)\,\sigma_V^{q^\prime}(z_2)
        - 2 m_{H_2}^2\,\tau^2\,\sigma_V^u(z_1)\,
                \frac{d}{dz_2}\sigma_V^{q^\prime}(z_2)\right]
\end{eqnarray}
and 
\begin{eqnarray}
z_1 & = & u^2-2 u \,\nu \,E\,, \\
z_2 & = & u^2 - 2 u \,\nu\, (E- X) - m_{H_2}^2 + 
2 i \,m_{H_2} \,\gamma \,u \,\sqrt{1-\nu^2} \,,\\
\label{defX}
X & = & (m_{H_1}/2)\,[1 + (m_{H_2}^2-t)/m_{H_1}^2]\,,\\
\tau & = & u\,\sqrt{1-\nu^2}\,\sqrt{1-\gamma^2}\,.
\end{eqnarray}
We note that because we have assumed isospin symmetry $\sigma^u$ also
represents a $d$-quark and, to illustrate Eq.~(\ref{fphl}), the $B^0\to \pi^-
\ell^+ \nu_\ell$ decay is characterised by
\begin{equation}
f_+^{B\pi}(t)= \kappa_d\frac{\surd 2}{f_\pi}
                \frac{N_c}{32\pi^2}\,
                F_{d}(t;E,m_B,m_\pi) \,.
\end{equation}

\subsection{Leptonic decays of a heavy meson} 
We are also interested in the leptonic decay of a heavy, pseudoscalar meson,
which is described by the matrix element
\begin{eqnarray}
\lefteqn{\langle 0 | \bar q \gamma_\mu\gamma_5 Q| P_{H_1}(p)\rangle := }\\
\nonumber & & 
f_{H_1} p_\mu = 
\frac{N_c}{(2\pi)^4}\int d^4k\,
        {\rm tr}\left[\gamma_5\gamma_\mu S_Q(k+p)
                \Gamma_{H_1}(k;p)S_q(k)\right]\,,
\end{eqnarray}
where $f_{H_1}$ is a single, dimensioned constant whose value describes all
strong interaction contributions to this weak decay.  For light mesons it has
been studied extensively~\cite{dserev,mr97} and with this normalisation
$f_\pi=0.131\,$GeV.  Using Eqs.~(\ref{genS}), (\ref{hqf}), (\ref{hmbsa}) and
(\ref{bsanormH}) one obtains an expression for $f_{H_1}$ valid in the
heavy-quark limit~\cite{mishaA}:
\begin{equation}
\label{fheavy}
f_{H_1} = \frac{\kappa_f}{\surd m_{H_1}}
\frac{N_c}{8\pi^2}\,\int_0^\infty\,du\,(\surd u - E)\,
\varphi(z)\,\left[\sigma_S^f(z) + \case{1}{2}\surd u \, \sigma_V^f(z)
\right]\,,
\end{equation}
where $z=u - 2 E \surd u$.  It follows that in the heavy-quark limit
\begin{equation}
\label{fscaling}
f_{H_f} \propto \frac{1}{\surd m_{H_f}}\,.
\end{equation}

This scaling law is counter to the trend observed in the light mesons, as
highlighted in Ref.~\cite{pmhl}, where $f_H$ increases at least up to
current-quark masses three-times that of the $s$-quark.  Contemporary
estimates of $f_D$ and $f_B$, such as those analysed in Ref.~\cite{hqlat},
suggest that Eq.~(\ref{fscaling}) is also not obeyed by experimentally
accessible heavy mesons.  The determination of the current-quark mass at
which the light meson trend is reversed, and that at which this heavy-quark
scaling law is satisfied, is an interesting, open question.

\section{Results and Discussion}
We have now defined all that is necessary for our calculation of the
semileptonic heavy $\to$ heavy and heavy $\to $ light meson transition form
factors and heavy-meson leptonic decay constants.  We have two free
parameters: the binding energy, $E$, introduced in Eq.~(\ref{vel}) and the
width, $\Lambda$, of the heavy meson Bethe-Salpeter amplitude, introduced in
Eq.~(\ref{phia}).  The dressed light-quark propagators and light-meson
Bethe-Salpeter amplitudes have been fixed completely in the application of
this framework to the study of $\pi$- and $K$-meson properties.

Our primary goal is to determine whether, with these two parameters, a
description and correlation of existing data is possible using the DSE
framework.  This was certainly true in our analysis of heavy $\to$ heavy
transitions alone~\cite{mishaA}.  We found that the function $\xi(w)$
necessarily has significant curvature and that a linear fit on $1\leq w\leq
1.6$ is inconsistent with our study.  However, our calculated value of the
slope parameter
\begin{equation}
\rho^2 := - \left.\frac{d}{dw}\xi(w)\right|_{w=1}
\end{equation}
was too strongly influenced by the experimental fit to the $B\to D$ data for
that study to provide an independent prediction of $\rho^2$.\footnote{In our
framework the minimum possible value for $\rho^2$ is
$1/3$~\protect\cite{mishaA}.}  Herein we eliminate this bias by excluding
$D$-meson observables from our primary procedure for fitting $E$ and
$\Lambda$.  This also facilitates an elucidation of where $1/\hat
M_c$-corrections are important.

\begin{figure}
\centering{\
\epsfig{figure=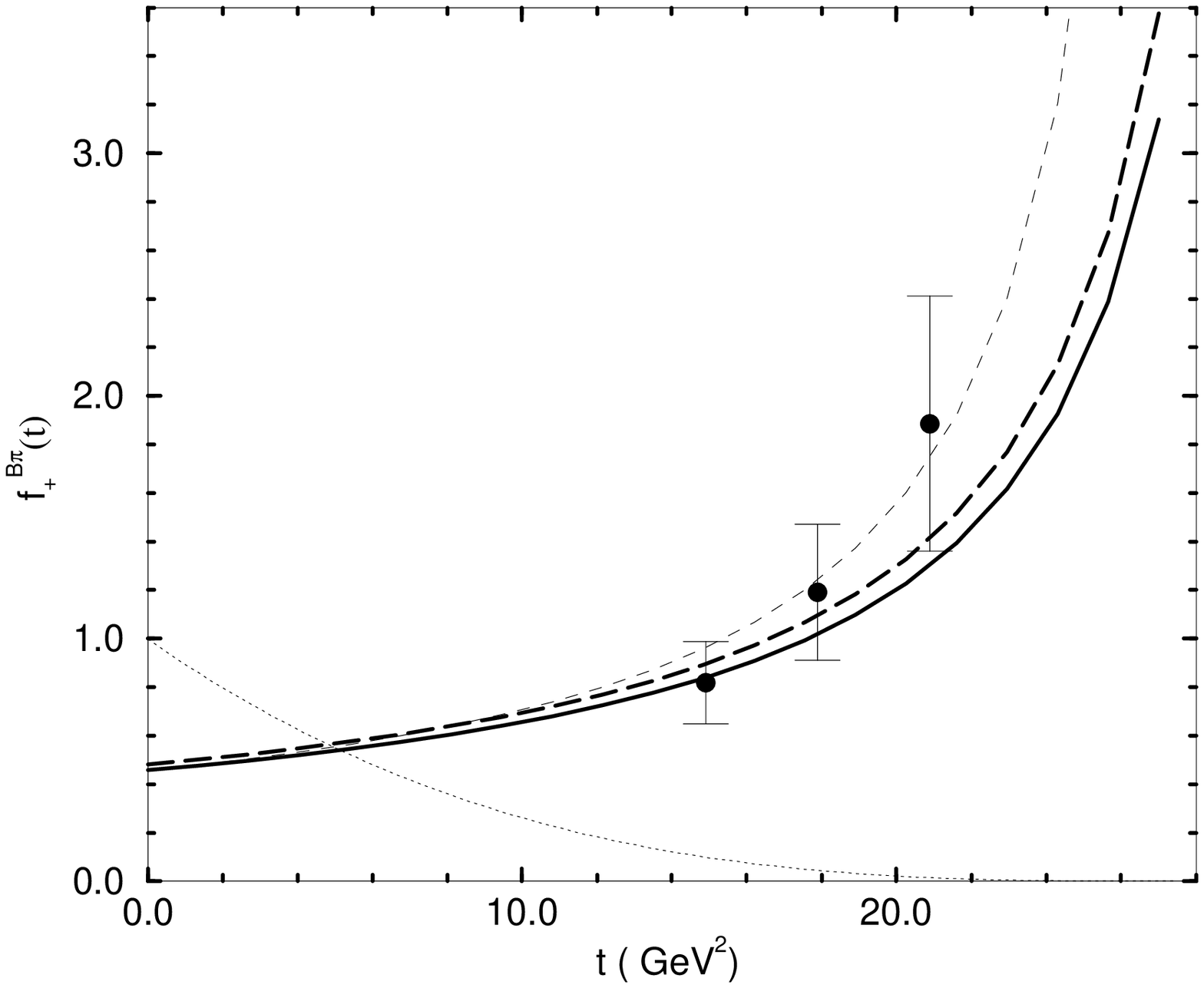,height=13.0cm}}
\caption{Our calculated form of $f_+^{B\pi}(t)$: solid line - column~1,
Table~\protect\ref{tablea}; dashed line - column~2,
Table~\protect\ref{tablea}.  For comparison, the data are the results of a
lattice simulation~\protect\cite{latt} and the light, short-dashed line is a
vector dominance, monopole model: $f_+(t)= 0.46/(1-t/m_{B^\ast}^2)$,
$m_{B^\ast} = 5.325\,$GeV.  The light, dotted line is the phase space factor
$|f_+^{B\pi}(0)|^2 \left[(t_+-t)(t_--t)\right]^{3/2}/(\pi m_B)^3$ in
Eq.~(\protect\ref{branching}), which illustrates that the $B\to \pi e \nu$
branching ratio is determined primarily by the small-$q^2$ behaviour of this
form factor.
\label{figbpi}}
\end{figure}

Our key results are presented in column one of Table~\ref{tablea}.  In
obtaining these results we varied $E$ and $\Lambda$ in order to obtain a
best, weighted least-squares fit to the three available lattice data
points~\cite{latt} for $f_+^{B\pi}$ and the experimental value~\cite{cleo96}
for the ${ B}^0\to\pi^- \ell^+\nu$ branching ratio.  In doing this we
constrained our study to yield $f_B = 0.17\,$GeV from Eq.~(\ref{fheavy}),
which is the central value favoured in a recent analysis of lattice
simulations~\cite{hqlat}, and used $m_B=5.27\,$GeV.  This fitting procedure
assumes only that the $b$-quark is in the heavy-quark domain; i.e., that
$1/\hat M_b$-corrections to the formulae we have derived herein are
negligible.  Our calculated form of $f_+^{B\pi}(t)$ is presented in
Fig.~\ref{figbpi}.  A good {\it interpolation} of our result is provided by
\begin{equation}
f_+^{B\pi}(t)= \frac{0.458}{1 - t/m_{\rm mon}^2}\,,
\; m_{\rm mon} = 5.67\,{\rm GeV} \,.
\end{equation}
This value of $m_{\rm mon}$ can be compared with that obtained in a fit to
lattice data~\cite{latt}: $m_{\rm mon}= 5.6 \pm 0.3$.

In Table~\ref{comp} we compare our favoured, calculated value of
$f_+^{B\pi}(0)= 0.46$ with this quantity obtained using a range of other
theoretical tools.  Since the $t$-dependence of $f_+^{B\pi}(t)$ is an outcome
of our calculation, the value we predict for $f_+^{B\pi}(0)$ is the only one
that allows simultaneous agreement between our calculations and existing
results of lattice simulations and the measured branching ratio.  If these
data are correct then in our framework it is not possible to obtain a value
of $f_+^{B\pi}(0)$ that differs from this favoured value by more than 10\%
unless the calculated $t$-dependence is changed significantly.  This could
only be effected by a modification of the vertex Ansatz, Eq.~(\ref{bare}),
and hence the accuracy of our prediction can be seen as a test of the
veracity of this Ansatz in the heavy-quark limit.

\begin{figure}
\centering{\
\epsfig{figure=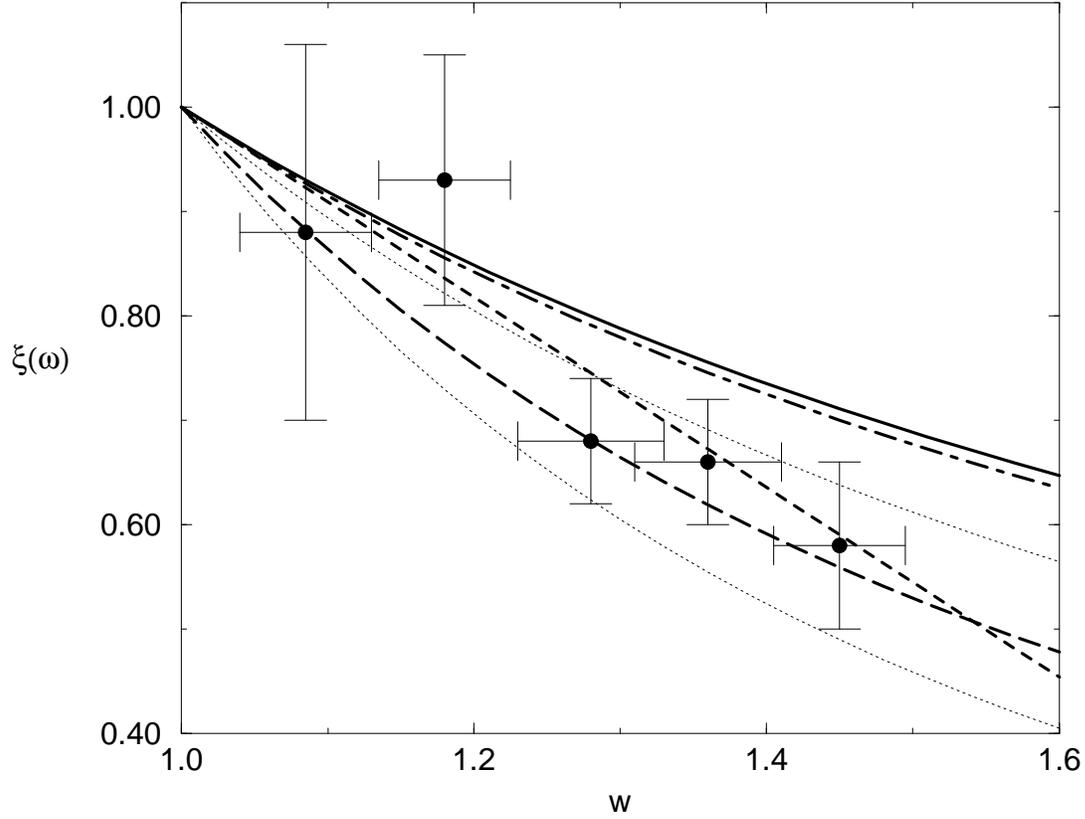,height=13.0cm}}
\caption{A comparison of our calculated form of $\xi(w)$ with recent
experimental analyses.  Our results: solid line - column~1,
Table~\protect\ref{tablea}; dot-dashed line - column~2,
Table~\protect\ref{tablea}. Experiment: data points -
Ref.~\protect\cite{argus93}; short-dashed line - linear fit from
Ref.~\protect\cite{cesr96}, see our Eq.~(\protect\ref{cesrlinear});
long-dashed line - nonlinear fit from Ref.~\protect\cite{cesr96}, see our
Eq.~(\protect\ref{cesrnonlinear}).  The two light, dotted lines are this
nonlinear fit evaluated with the extreme values of $\rho^2$: upper line,
$\rho^2= 1.17$ and lower line, $\rho^2=1.89$.
\label{figiwfn}}
\end{figure}

In Fig.~\ref{figiwfn} we present our calculated form of the function,
$\xi(w)$, that characterises the semileptonic heavy $\to$ heavy meson decays.
We have compared our calculation with the experimental results of
Ref.~\cite{argus93} and the following fits to the experimental data in
Ref.~\cite{cesr96}:
\begin{eqnarray}
\xi(w) & = & 1 - \rho^2\,( w - 1), \; \rho^2 = 0.91\pm 0.15 \pm 0.16\,,
\label{cesrlinear}\\
\xi(w) & = & \frac{2}{w+1}\,\exp\left[(1-2\rho^2) \frac{w-1}{w+1}\right],
        \;\rho^2 = 1.53 \pm 0.36 \pm 0.14\,.
\label{cesrnonlinear}
\end{eqnarray}
Our calculated result for $\rho^2$ is close to that in Eq.~(\ref{cesrlinear})
but our form of $\xi(w)$ has significant curvature and deviates quickly from
the linear fit.  The curvature is, in fact, very well matched to that of the
fit in Eq.~(\ref{cesrnonlinear}), however, the value of $\rho^2$ listed in
that case is very different to our calculated value.  

In Ref.~\cite{mishaA} we fitted $E$ and $\Lambda$ to the nonlinear form in
Eq.~(\ref{cesrnonlinear}) and fitted it exactly.  We believe that part of the
discrepancy observed here is due to our neglect of $1/\hat M_c$-corrections
in the calculation of $\xi(w)$, the magnitude of which is exposed because of
our newfound ability to constrain our parameters without referring to
$D$-meson observables.  Nevertheless, the agreement between this calculation
and the data is reasonable, with the difference largest at $\omega_{\rm max}$
where it is a little more than one standard deviation.  Hence $1/\hat
M_c$-corrections cannot be too large.

\begin{figure}
\centering{\
\epsfig{figure=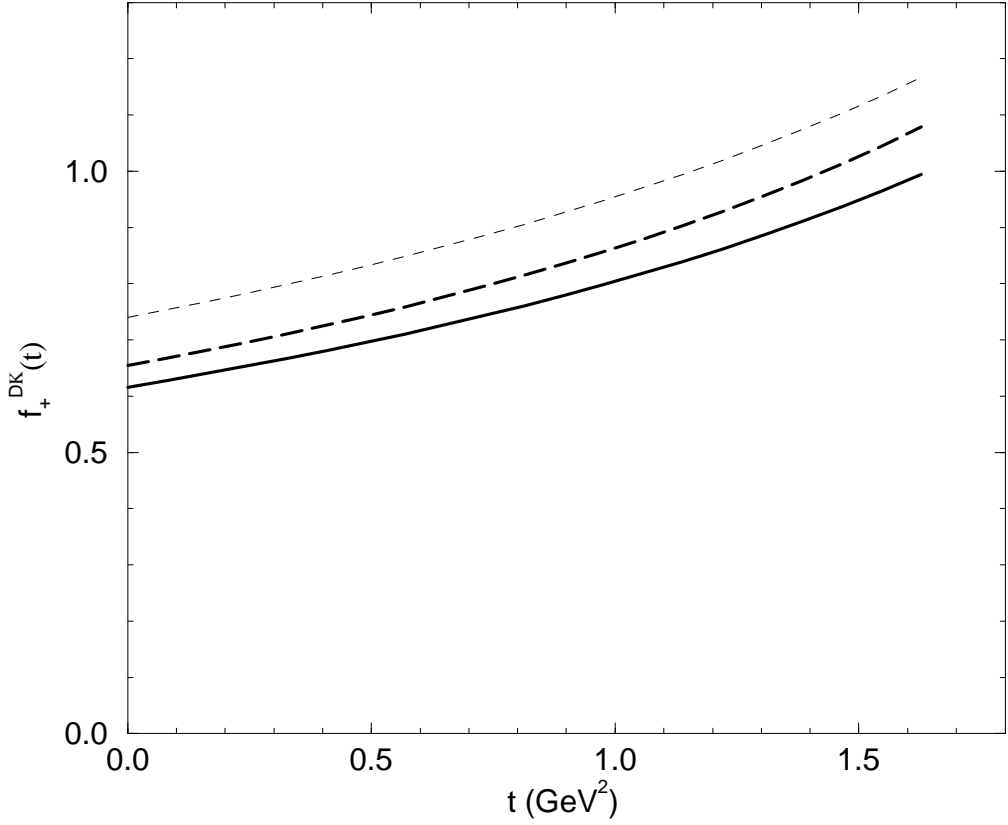,height=13.0cm}}
\caption{Our calculated form of $f_+^{DK}(q^2)$: solid line - column~1,
Table~\protect\ref{tablea}; dashed line - column~2,
Table~\protect\ref{tablea}.  For comparison the light, short-dashed line is
a vector dominance, monopole model: $f_+(q^2)= 0.74/(1-q^2/m_{D_s^\ast}^2)$,
$m_{D_s^\ast} = 2.11\,$GeV.  
\label{figdk}}
\end{figure}

In Fig.~\ref{figdk} we present our calculated form of $f_+^{DK}(t)$.  The
$t$-dependence is well-approximated by a monopole fit.  Our favoured,
calculated value of $f_+^{DK}(0) = 0.62$ is approximately 15\% less than the
experimental value~\cite{pdg96}.  We interpret this as a {\it gauge} of the
size of $1/\hat M_c$-corrections.  These corrections are expected to reduce
the value of the $D$-meson leptonic decay constants from that obtained using
Eq.~(\ref{fheavy}).  A 15\% reduction in the $D$-meson leptonic decay
constants in column one of Table~\ref{tablea} yields $f_D = 0.24\,$GeV and
$f_{D_s}=0.26\,$GeV, values which are consistent with lattice
estimates~\cite{hqlat} and the latter with experiment~\cite{expfds}.

We have also calculated $f_+^{D\pi}(t)$ and find that on the kinematically
accessible domain, $0 < t < (m_D - m_\pi)^2$, the following monopole form
provides an excellent {\it interpolation}
\begin{equation}
f_+^{D\pi}(t) = \frac{0.716}{1 - t/m_{\rm mon}^2}\,,
\; m_{\rm mon} = 2.15\,{\rm GeV} \,.
\end{equation}
We note that a naive vector meson dominance assumption would lead one to
expect $m_{\rm mon} \approx m_{D^*}= 2.0\,{\rm GeV}\,$.  Using $(E,\Lambda)$
from Table~\ref{tablea} we obtain
\begin{equation}
R_\pi := \frac{Br(D \to \pi \ell \nu)}{Br(D\to K \ell \nu)}
= 2.47 \left|\frac{V_{cd}}{V_{cs}}\right|^2 = 0.13\,,
\end{equation}
for $|V_{cd}/V_{cs}|^2 = 0.051 \pm 0.002$~\cite{pdg96}, and in this ratio the
bulk of the $1/\hat M_c$-corrections should cancel.  Experimentally
\begin{eqnarray}
\label{rpia}
R_\pi = \frac{Br(D^0\to \pi^- e^+ \nu_e)}{Br(D^0\to  K^- e^+ \nu_e)}
        & = & 0.11 ^{+0.06}_{-0.03} \pm 0.1~\protect\cite{pdg96,mark3}\,,\\
\label{rpib}
R_\pi = 2 \frac{Br(D^+\to \pi^0 e^+ \nu_e)}{Br(D^+\to  \bar K^0 e^+ \nu_e)}
        & = & 0.17 \pm 0.05 \pm 0.03~\protect\cite{cleoii}\,.
\end{eqnarray}
We observe that if one makes the assumption of single-pole, $D^*$ and $D_s^*$
vector meson dominance for the $t$-dependence of the form factors
$f_+^{D\pi}$ and $f_+^{DK}$, respectively, one obtains the simple formula
\begin{equation}
R_\pi= 1.97 \left|\frac{f_+^{D\pi}(0)}{f_+^{DK}(0)}\right|^2
                \left|\frac{V_{cd}}{V_{cs}}\right|^2\,.
\end{equation}
This approach has been employed~\cite{pdg96} in order to estimate
$f_+^{D\pi}(0)/f_+^{DK}(0) = 1.0^{+0.3}_{-0.2}\pm 0.04$ or $1.3\pm0.2\pm 0.1$
from Eqs.~(\ref{rpia}) and (\ref{rpib}).  We calculate
\begin{equation}
\frac{f_+^{D\pi}(0)}{f_+^{DK}(0)} = 1.16\,.
\end{equation}

It is incumbent upon us now to stress that we explicitly {\it do not}\ assume
vector meson dominance.  Our calculated results reflect only the importance
and influence of the dressed-quark and -gluon substructure of the heavy
mesons.  This substructure is manifest in the dressed propagators and bound
state amplitudes, which fully determine the value of every quantity
calculated herein.  Explicit vector meson contributions would appear as pole
terms in ${\cal V}_\mu^{f_1f_2}(k_1,k_2)$, which are excluded in our Ansatz,
Eq.~(\ref{bare}).  That simple-pole Ans\"atze provide efficacious
interpolations of our results on the accessible kinematic domain is not
surprising, given that the form factor must rise slowly away from its value
at $t=0$ and the heavy meson mass provides a dominant intrinsic scale, which
is modified slightly by the scale in the light-quark propagators and meson
bound state amplitudes.  Similar observations are true in the calculation of
the pion form factor, as discussed in detail in Sec.~7.1 of
Ref.~\cite{pctrev} and Sec. 2.3.1 of Ref.~\cite{cdrpion}.

In column two of Table~\ref{tablea} we present the results obtained when $E$
and $\Lambda$ are varied in order to obtain a best, weighted least-squares
fit to: the lattice data on $f_+^{B\pi}$; the ${ B}^0\to\pi^- \ell^+\nu$
branching ratio; and the experimental data on $\xi(w)$ reported in
Ref.~\cite{argus93}.  The latter introduce $D$-meson properties into our
fitting constraints but their effect on our calculations is not very
significant.  The tabulated quantity most affected is the ${ B}^0\to\pi^-
\ell^+\nu$ branching ratio but this increases by only 15\% and remains
acceptably close to the experimental value.  The effect that this modified
fitting procedure has on the transition form factors is also small, as
illustrated by the comparisons in Figs.~\ref{figbpi}-\ref{figdk}.  Not
surprisingly, the largest effect is a uniform 5\% increase in the magnitude
of $f_+^{DK}(t)$.

In Table~\ref{tabled} we present the results obtained using the different
functional form for the heavy-meson Bethe-Salpeter amplitude in
Eq.~(\ref{phib}).  A direct comparison with the results in Table~\ref{tablea}
indicates that our results are insensitive to such details and hence are
robust.  The binding energy, $E$, is unchanged and the width,
$\tilde\Lambda$, is smaller, as expected since Eq.~(\ref{phib}) does not
decrease as rapidly with $k^2$ as the form in Eq.~(\ref{phia}).  A
quantitative statement of this is that
\begin{eqnarray}
\int_0^\infty\,dk^2\,\left({\rm e}^{-k^2/\Lambda^2}\right)^2
& = & \case{1}{2}\,\Lambda^2\,, \\
\int_0^\infty\,dk^2\,\left(\frac{\tilde\Lambda^2}{k^2+\tilde\Lambda^2}\right)^2
& = & \tilde\Lambda^2
\end{eqnarray}
and $\tilde\Lambda = 0.92\,$GeV $\sim \Lambda /\surd 2 = 1.0\,$GeV is just
that reduction necessary to provide the same integrated strength for both
amplitudes.

Tables~\ref{tableb} and \ref{tablec} provide a further elucidation of the
impact of possible systematic errors in our calculation.  These results are
obtained through a repetition of the calculations that yield
Table~\ref{tablea} but with $f_B$ constrained to be $0.135$ and $0.205\,$GeV,
respectively, which are the outer limits estimated in an analysis of
contemporary lattice simulations~\cite{hqlat}.  In the direct application of
the methods of Ref.~\cite{mr97} to heavy mesons the value of $f_B$ would be a
prediction.  Herein, since we do not calculate but instead fit the
heavy-meson Bethe-Salpeter amplitude, $f_B$ acts as a constraint on the
width, $\Lambda$, of the Bethe-Salpeter amplitude, as seen in a comparison of
Tables~\ref{tablea}, \ref{tableb} and \ref{tablec}.  The binding energy, $E$,
is then the only true free parameter and it varies over a range of no more
than 8\%.  Comparing these tables, we see that our results are not very
sensitive to the value of $f_B$ in the range we have explored; i.e., our
results are robust.

We judge that the best description of the available data is obtained with
$f_B=0.17\,$GeV, with a lower value, $f_B \to 0.135\,$GeV, more acceptable
than a higher one.  The value of $E=0.44\,$GeV that provides this best
description can be compared with the value of $E_{\rm bind}\sim
0.25$-$0.35\,$GeV obtained in a lattice NRQCD simulation~\cite{alikhan}.  The
value of $\tilde\Lambda =0.92\,$GeV indicates that the heavy meson occupies a
spacetime volume only 15\% of that occupied by the pion.

\section{Conclusions}
Using the same phenomenological Dyson-Schwinger equation (DSE) framework
employed in successful studies of light meson observables as diverse as
$\pi$-$\pi$ scattering~\cite{pipi} and diffractive electroproduction of
vector mesons~\cite{pich97}, we have analysed semileptonic heavy $\to$ heavy
and heavy $\to$ light meson transition form factors.  In this application we
introduced and explored a heavy-quark limit of the DSEs based on the
observation that the mass function of heavy quarks evolves slowly with
momentum.

With two parameters: $E$, the difference between the heavy-meson mass and the
effective-mass of the heavy quark; and $\Lambda$, the width of the heavy
meson Bethe-Salpeter amplitude, we obtained a uniformly good, robust
description of all available $B\to\pi$ data with a prediction for
$f_+^{B\pi}(t)$ on the kinematically accessible $t$-domain.  In analysing
$B\to D$, $D\to K$ and $D\to \pi$ transitions we estimated that
$1/m_c$-corrections to our heavy-quark limit contribute no more than 15\%.  A
significant feature of our study is the correlation of heavy $\to$ heavy and
heavy $\to$ light transitions {\it and} their correlation with light meson
observables, which are dominated by effects such as dynamical chiral symmetry
breaking and confinement.

This study can be extended, with the application of the framework to
semileptonic decays with vector meson final states using no additional
parameters.  It can also be improved, for example, by an exploration of the
effect of more sophisticated Ans\"atze for the dressed-quark-W-boson vertex
and of the inclusion of all amplitudes in the light-meson Bethe-Salpeter
amplitude with refitted light-quark propagators.

A more significant qualitative improvement is the direct study of the
Bethe-Salpeter equation for heavy mesons using the methods of
Ref.~\cite{mr97}; Ref.~\cite{pmhl} is the pilot.  This programme involves the
important step of critically analysing the reliability for heavy quarks of
ladder-like truncations of the dressed-quark-antiquark scattering kernel in
both the quark DSE and meson Bethe-Salpeter equation.  Addressing this
question and developing an efficacious truncation will allow a {\it
correlation} of heavy- and light-meson observables via the few parameters
that characterise the behaviour of the quark-quark interaction in the
nonperturbative domain; i.e., relate both heavy- and light-meson observables
to the long-range part of the quark-quark interaction.

\acknowledgements MAI gratefully acknowledges the hospitality of the Physics
Division at ANL and CDR that of the BLTP and the LCTP at the JINR during
visits where some of this work was conducted.  This work was supported in
part by the US Department of Energy, Nuclear Physics Division, under contract
number W-31-109-ENG-38 and benefited from the resources of the National
Energy Research Scientific Computing Center.

\appendix
\section*{A Derivation}
A typical integral arising in the detailed analysis of Eq.~(\ref{ia}) has the
form 
\begin{equation}
\label{typical}
J= \int \frac{d^4 k}{\pi^2}\, \frac{1}{k\cdot v - E}
\,Z(k^2)\,\sigma([k-p_2]^2)\,. 
\end{equation}
To simplify it we introduce a Laplace transform for the functions $Z(k^2)$
and $\sigma([k-p_2]^2)$:
\begin{eqnarray}
Z(k^2) & = &\int_0^\infty ds\,\tilde Z (s) {\rm e}^{-s k^2}\,, \\
\sigma([k-p_2]^2)& = &
\int_0^\infty du\,\tilde \sigma (u) {\rm e}^{-u [k-p_2]^2} 
\end{eqnarray}
and a Gaussian representation of the heavy-quark propagator:
\begin{equation}
\frac{1}{k\cdot v - E} = \int_0^\infty d\alpha\,
        {\rm e}^{-\alpha (k\cdot v - E)}\,.
\end{equation}
Inserting these identities we obtain
\begin{eqnarray}
J & = &\int\limits_0^\infty ds \,\tilde Z(s)
\int\limits_0^\infty du \,\tilde \sigma(u)
\int\limits_0^\infty d\alpha\,
\int \frac{d^4 k}{\pi^2}\,
\exp\{-sk^2-\alpha (k\cdot v- E)-u [k+p_2]^2\} \\
& = & 
\int\limits_0^\infty ds \,\tilde Z(s)
\int\limits_0^\infty du \,\tilde \sigma(u)
\int\limits_0^\infty d\alpha\,
\exp\left\{\alpha E -u p_2^2 + (u p_2 + \case{1}{2}\alpha v)^2/(s+u)\right\}\\
& & \nonumber
        \int \frac{d^4 k}{\pi^2}\,
        \exp\left\{-(s+u)
        \left[k + (u p_2 + \case{1}{2}\alpha v)/(s+u)\right]^2\right\}\,.
\end{eqnarray}
Shifting variables: $k \to k - (u p_2 + \case{1}{2}\alpha v)/(s+u)$ and
subsequently $\alpha \to (s+u)\alpha$, yields
\begin{eqnarray}
\lefteqn{J  = \int\limits_0^\infty ds \,\tilde Z(s)
\int\limits_0^\infty du \,\tilde \sigma(u)}\\
& & \nonumber
\int\limits_0^\infty d\alpha\,(s+u) \,
\exp\left\{ - (s+u) (\case{1}{4}\alpha^2 - \alpha E) + u \alpha X 
        - \frac{s u }{s+u}\,p_2^2 \right\} 
        \int \frac{d^4 k}{\pi^2}\,{\rm e}^{-(s+u)k^2}\\
& = & \int\limits_0^\infty ds \,\tilde Z(s)
\int\limits_0^\infty du \,\tilde \sigma(u)
\int\limits_0^\infty d\alpha\,\frac{1}{s+u}
\exp\left\{ - (s+u) (\case{1}{4}\alpha^2 - \alpha E) - u \alpha X 
        -\frac{su}{s+u}\,p_2^2 \right\}
\end{eqnarray}
where $X:= -v\cdot p_2$, Eq.~(\ref{defX}).  Making use of the identities
\begin{eqnarray}
\exp\left\{-\frac{s u}{s+u}\,p_2^2\right\}
& = & \sqrt\frac{s + u}{\pi}\,\int_{-\infty}^\infty\,
        d\tau\,\exp\left\{-s\tau^2 
        - u \left(\tau + \sqrt{p_2^2}\right)^2\right\}\,,\\
\frac{1}{\sqrt{s+u}}
& = & \frac{1}{\surd \pi}\,\int_{-\infty}^\infty\,dt\,
        {\rm e}^{-(s+u)t^2}
\end{eqnarray}
we obtain
\begin{eqnarray}
\lefteqn{J = \frac{2}{\pi}\int_0^\infty\,d\alpha\,
        \int_{-\infty}^{\infty}\,d\tau}\\
& & \nonumber
        \int_{-\infty}^{\infty}\,dt\,
        Z(\alpha^2 -2 \alpha E + \tau^2 + t^2 )\,
        \sigma\left(\alpha^2 -2 \alpha E + 2 \alpha X 
                + \left(\tau + \sqrt{p_2}\right)^2 + t^2 \right)\,.
\end{eqnarray}
Introducing spherical polar coordinates
\begin{eqnarray}
\alpha & = & u\,\nu\,, \\
\tau & = & u\,\sqrt{1-\nu^2}\gamma\,, \\
t & = & u\,\sqrt{1-\nu^2}\,\sqrt{1-\gamma^2}\,,
\end{eqnarray}
with $u\in[0,\infty)$, $\nu\in [0,1]$ and $\gamma\in[-1,1]$, we arrive at
\begin{eqnarray}
J & = & \frac{4}{\pi}\int_{-1}^1\frac{d\gamma}{\sqrt{1-\gamma^2}}\,
        \int_0^1\,d\nu\,\int_0^\infty\,du\,u^2\,
        Z(z_1)\,\sigma(z_2)\,,
\end{eqnarray}
where, using $p_2^2 = - m_{H_2}^2$, 
\begin{eqnarray}
z_1 & = & u^2 - 2\,u\,\nu\,E\,, \\
z_2 & = & u^2 - 2 \,u\,\nu\,(E-X)\, - m_{H_2}^2
        + 2\,i\,m_{H_2}\,u\,\sqrt{1-\nu^2}\,\sqrt{1-\gamma^2}\,.
\end{eqnarray}
This is recognisably of the form in Eq.~(\ref{fphl}).

Structures more complicated than Eq.~(\ref{typical}) arise in deriving the
complete form of $F_{q^\prime}$, however, they can all be analysed and
simplified using analogues of the method illustrated above.


%
\begin{table}
\begin{tabular}{|l|l|l|l|}
 & DATA/ESTIMATES &    \multicolumn{2}{c|}{$f_B=0.170\,$GeV} \\\hline
$(E,\Lambda)\,$(GeV)  & & (0.442,1.408) & (0.465,1.405) \\\hline
$\Sigma^2/N $        & &  0.48        &  1.22\\\hline\hline
 $f_+^{B\pi}(14.9\,{\rm GeV}^2)$  &  $0.82 \pm 0.17$~\protect\cite{latt}& 
        0.84$^\dagger$ & 0.89$^\dagger$ \\\hline
 $f_+^{B\pi}(17.9\,{\rm GeV}^2)$   &  $1.19 \pm 0.28$~\protect\cite{latt} & 
        1.02$^\dagger$ & 1.09$^\dagger$ \\\hline
 $f_+^{B\pi}(20.9\,{\rm GeV}^2)$ &  $1.89 \pm 0.53$~\protect\cite{latt}  & 
        1.30$^\dagger$ & 1.41$^\dagger$ \\\hline
Br$({B}^0\to\pi^- \ell^+\nu)$  &  
  $[1.8\pm 0.4 \pm 0.3\pm 0.2 ]\times 10^{-4}$~\protect\cite{cleo96} & 
  2.0 $\times 10^{-4}$$^\dagger$  & 2.3$\times 10^{-4}$$^\dagger$  \\ \hline
 $f_+^{B\pi}(0)$  &  $0.18 \to 0.49$~\protect\cite{lellouch96}& 0.46 & 0.48 \\\hline
 $f_+^{DK}(0)$  &  0.74 $\pm$ 0.03~\protect\cite{pdg96}&  $0.62$ & 0.65\\\hline
$\xi(1.085\pm 0.045) $& $0.88 \pm 0.18$~\protect\cite{argus93} & 
        0.93 & 0.93$^\dagger$\\\hline
$\xi(1.18\pm 0.045) $& $0.93 \pm 0.12$~\protect\cite{argus93} & 
        0.86 & 0.86$^\dagger$\\\hline
$\xi(1.28\pm 0.050)$ & $0.68 \pm 0.06$~\protect\cite{argus93} & 
        0.80 & 0.79$^\dagger$\\\hline
$\xi(1.36\pm 0.050)$ & $0.66 \pm 0.06$~\protect\cite{argus93} & 
        0.76 & 0.75$^\dagger$\\\hline
$\xi(1.45\pm 0.045)$ & $0.58 \pm 0.08$~\protect\cite{argus93} & 
        0.71 & 0.70$^\dagger$\\\hline
$\rho^2 $    & $\displaystyle \begin{array}{l}
                                0.91 \pm 0.15 \pm 0.06 \\
                                1.53 \pm 0.36 \pm 0.14
                                \end{array}$~\cite{cesr96} 
                              & 0.87 & 0.92 \\\hline
$f_{B_s}\,$(GeV) & $0.195 \pm 0.035$~\cite{hqlat} & 0.184 & 0.184\\\hline
$f_{B_s}/f_B$ & $1.14 \pm 0.08 $~\cite{hqlat} & 1.083 & 1.082 \\\hline
$f_{D}\,$(GeV) & $0.200 \pm 0.030 $~\cite{hqlat} & 0.285 & 0.285 \\\hline
$f_{D_s}\,$(GeV) &  $ 0.220 \pm 0.030$~\cite{hqlat}  & 0.304 & 0.304 \\\hline
$f_{D_s}/f_D$ & $1.10\pm 0.06$~\cite{hqlat}& 1.066 & 1.066 \\\hline
\end{tabular}
\caption{A comparison of our calculated results with available data when we
require $f_B=0.170\,$GeV, which is the central value estimated in
Ref.~\protect\cite{hqlat}, and use Eq.~(\protect\ref{phia}).
In each column the quantities marked by $^\dagger$ are those used to
constrain the parameters $(E,\Lambda)$ by minimising
\mbox{$\Sigma^2 := \sum_{i=1}^{N}\,([y_i^{\rm calc}-y_i^{\rm data}]/
\sigma(y)_i^{\rm
data})^2$},
where $N$ is the number of data items used.  The results in the first column
assume that heavy-quark symmetry is valid for the $b$-quark but do not rely
on this being true for the $c$-quark.  We note that: 1) our values of $f_D$
and $f_{D_s}$ are obtained via Eq.~(\protect\ref{fscaling}) from $f_B$ and
$f_{B_s}$, respectively, using $m_B=5.27$, $m_{B_s}=5.375$, $m_D= 1.87$ and
$m_{D_s}= 1.97\,$GeV; 2) the experimental determination of $\rho^2$ is
sensitive to the form of the fitting function, e.g., see
Ref.~\protect\cite{cesr96}; 3) an analysis of four experimental measurements
of $D_s\to\mu\nu$ decays yields $f_{D_s}=0.241 \pm 0.21 \pm
0.30\,$GeV~\protect\cite{expfds}.
\label{tablea}}
\end{table}
\begin{table}
\begin{tabular}{l|l}
 Reference              & $f_+^{B\pi}(0) $    \\\hline
Our Result              & 0.46  \\\hline
Dispersion relations~\cite{lellouch96} & 0.18 $\to$ 0.49 \\
Quark Model~\cite{qma}  & $ 0.33 \pm 0.06 $ \\
Quark Model~\cite{qmb}  & $0.21 \pm 0.02 $ \\
Quark Model~\cite{qmc}  & 0.29 \\
Light-Cone Sum Rules~\cite{lcsr} & $\!\!\left\{\begin{array}{l}
                                   0.29~{\rm direct}\\  
                                   0.44~{\rm pole~dominance}
                                        \end{array} \right.$\\
Quark Confinement Model~\cite{mishaB}
                        & 0.6 \\
Quark Confinement Model~\cite{mishaC,mishaD} & 0.53 
\end{tabular}
\caption{A comparison of our favoured, calculated result for $f_+^{B\pi}(0)$
with a representative but not exhaustive list of values obtained using other
theoretical tools.  More extensive and complementary lists are presented in
Refs.~\protect\cite{lellouch96,qmc,mishaD}.
\label{comp}}
\end{table}
\begin{table}
\begin{tabular}{|l|l|l|}
 & DATA/ESTIMATES &    $f_B=0.170\,$GeV \\\hline
$(E,\Lambda)\,$(GeV)  & & (0.455,0.918)  \\\hline
$\Sigma^2/N $        & &  0.46        \\\hline\hline
 $f_+^{B\pi}(14.9\,{\rm GeV}^2)$  &  $0.82 \pm 0.17$~\protect\cite{latt}& 
        0.84$^\dagger$ \\\hline
 $f_+^{B\pi}(17.9\,{\rm GeV}^2)$   &  $1.19 \pm 0.28$~\protect\cite{latt} & 
        1.02$^\dagger$ \\\hline
 $f_+^{B\pi}(20.9\,{\rm GeV}^2)$ &  $1.89 \pm 0.53$~\protect\cite{latt}  & 
        1.32$^\dagger$ \\\hline
Br$({B}^0\to\pi^- \ell^+\nu)$  &  
  $[1.8\pm 0.4 \pm 0.3\pm 0.2 ]\times 10^{-4}$~\protect\cite{cleo96} & 
  2.0 $\times 10^{-4}$$^\dagger$   \\ \hline
 $f_+^{B\pi}(0)$  &  $0.18 \to 0.49$~\protect\cite{lellouch96}& 0.45 \\\hline
 $f_+^{DK}(0)$  &  0.74 $\pm$ 0.03~\protect\cite{pdg96}&  $0.62$ \\\hline
$\xi(1.085\pm 0.045) $& $0.88 \pm 0.18$~\protect\cite{argus93} & 
        0.92 \\\hline
$\xi(1.18\pm 0.045) $& $0.93 \pm 0.12$~\protect\cite{argus93} & 
        0.84\\\hline
$\xi(1.28\pm 0.050)$ & $0.68 \pm 0.06$~\protect\cite{argus93} & 
        0.77 \\\hline
$\xi(1.36\pm 0.050)$ & $0.66 \pm 0.06$~\protect\cite{argus93} & 
        0.72 \\\hline
$\xi(1.45\pm 0.045)$ & $0.58 \pm 0.08$~\protect\cite{argus93} & 
        0.67\\\hline
$\rho^2 $    & $\displaystyle \begin{array}{l}
                                0.91 \pm 0.15 \pm 0.06 \\
                                1.53 \pm 0.36 \pm 0.14
                                \end{array}$~\cite{cesr96} 
                              & 1.03\\\hline
$f_{B_s}\,$(GeV) & $0.195 \pm 0.035$~\cite{hqlat} & 0.180 \\\hline
$f_{B_s}/f_B$ & $1.14 \pm 0.08 $~\cite{hqlat} & 1.061 \\\hline
$f_{D}\,$(GeV) & $0.200 \pm 0.030 $~\cite{hqlat} & 0.285\\\hline
$f_{D_s}\,$(GeV) &  $ 0.220 \pm 0.030$~\cite{hqlat}  & 0.298 \\\hline
$f_{D_s}/f_D$ & $1.10\pm 0.06$~\cite{hqlat}& 1.044 \\\hline
\end{tabular}
\caption{A comparison of our calculated results with available data when we
require $f_B=0.170\,$GeV, which is the central value estimated in
Ref.~\protect\cite{hqlat}, and use Eq.~(\protect\ref{phib}) .  (See
Table.~\protect\ref{tablea} for additional remarks and an explanation of the
symbols.)
\label{tabled}}
\end{table}
\begin{table}
\begin{tabular}{|l|l|l|l|}
 & DATA/ESTIMATES &    \multicolumn{2}{c|}{$f_B=0.135\,$GeV} \\\hline
$(E,\Lambda)\,$(GeV)  & & (0.457,1.138) & (0.466,1.135) \\\hline
$\Sigma^2/N $        & &  0.50        &  0.97\\\hline\hline
 $f_+^{B\pi}(14.9\,{\rm GeV}^2)$  &  $0.82 \pm 0.17$~\protect\cite{latt}& 
        0.86$^\dagger$ & 0.88$^\dagger$ \\\hline
 $f_+^{B\pi}(17.9\,{\rm GeV}^2)$   &  $1.19 \pm 0.28$~\protect\cite{latt} & 
        1.05$^\dagger$ & 1.08$^\dagger$ \\\hline
 $f_+^{B\pi}(20.9\,{\rm GeV}^2)$ &  $1.89 \pm 0.53$~\protect\cite{latt}  & 
        1.36$^\dagger$ & 1.40$^\dagger$ \\\hline
Br$({B}^0\to\pi^- \ell^+\nu)$  &  
  $[1.8\pm 0.4 \pm 0.3\pm 0.2 ]\times 10^{-4}$~\protect\cite{cleo96} & 
  2.1 $\times 10^{-4}$$^\dagger$  & 2.2$\times 10^{-4}$$^\dagger$  \\ \hline
 $f_+^{B\pi}(0)$  &  $0.18 \to 0.49$~\protect\cite{lellouch96}& 0.46 & 0.47 \\\hline
 $f_+^{DK}(0)$  &  0.74 $\pm$ 0.03~\protect\cite{pdg96}&  $0.64$ & 0.65\\\hline
$\xi(1.085\pm 0.045) $& $0.88 \pm 0.18$~\protect\cite{argus93} & 
        0.92 & 0.92$^\dagger$\\\hline
$\xi(1.18\pm 0.045) $& $0.93 \pm 0.12$~\protect\cite{argus93} & 
        0.85 & 0.85$^\dagger$\\\hline
$\xi(1.28\pm 0.050)$ & $0.68 \pm 0.06$~\protect\cite{argus93} & 
        0.78 & 0.78$^\dagger$\\\hline
$\xi(1.36\pm 0.050)$ & $0.66 \pm 0.06$~\protect\cite{argus93} & 
        0.74 & 0.73$^\dagger$\\\hline
$\xi(1.45\pm 0.045)$ & $0.58 \pm 0.08$~\protect\cite{argus93} & 
        0.69 & 0.69$^\dagger$\\\hline
$\rho^2 $    & $\displaystyle \begin{array}{l}
                                0.91 \pm 0.15 \pm 0.06 \\
                                1.53 \pm 0.36 \pm 0.14
                                \end{array}$~\cite{cesr96} 
                              & 0.96 & 0.98 \\\hline
$f_{B_s}\,$(GeV) & $0.195 \pm 0.035$~\cite{hqlat} & 0.148 & 0.148\\\hline
$f_{B_s}/f_B$ & $1.14 \pm 0.08 $~\cite{hqlat} & 1.096 & 1.096 \\\hline
$f_{D}\,$(GeV) & $0.200 \pm 0.030 $~\cite{hqlat} & 0.227 & 0.227 \\\hline
$f_{D_s}\,$(GeV) &  $ 0.220 \pm 0.030$~\cite{hqlat}  & 0.244 & 0.244 \\\hline
$f_{D_s}/f_D$ & $1.10\pm 0.06$~\cite{hqlat}& 1.079 & 1.078 \\\hline
\end{tabular}
\caption{A comparison of our calculated results with available data when we
require $f_B=0.135\,$GeV, which is the lower bound estimated in
Ref.~\protect\cite{hqlat}, and use Eq.~(\protect\ref{phia}).  (See
Table.~\protect\ref{tablea} for additional remarks and an explanation of the
symbols.)
\label{tableb}}
\end{table}
\begin{table}
\begin{tabular}{|l|l|l|l|}
 & DATA/ESTIMATES &    \multicolumn{2}{c|}{$f_B=0.205\,$GeV} \\\hline
$(E,\Lambda)\,$(GeV)  & & (0.469,1.677) & (0.479,1.678) \\\hline
$\Sigma^2/N $        & &  0.83        &  1.45\\\hline\hline
 $f_+^{B\pi}(14.9\,{\rm GeV}^2)$  &  $0.82 \pm 0.17$~\protect\cite{latt}& 
        0.91$^\dagger$ & 0.94$^\dagger$ \\\hline
 $f_+^{B\pi}(17.9\,{\rm GeV}^2)$   &  $1.19 \pm 0.28$~\protect\cite{latt} & 
        1.11$^\dagger$ & 1.15$^\dagger$ \\\hline
 $f_+^{B\pi}(20.9\,{\rm GeV}^2)$ &  $1.89 \pm 0.53$~\protect\cite{latt}  & 
        1.43$^\dagger$ & 1.49$^\dagger$ \\\hline
Br$({B}^0\to\pi^- \ell^+\nu)$  &  
  $[1.8\pm 0.4 \pm 0.3\pm 0.2 ]\times 10^{-4}$~\protect\cite{cleo96} & 
  2.4 $\times 10^{-4}$$^\dagger$  & 2.5$\times 10^{-4}$$^\dagger$  \\ \hline
 $f_+^{B\pi}(0)$  &  $0.18 \to 0.49$~\protect\cite{lellouch96}& 0.49 & 0.50 \\\hline
 $f_+^{DK}(0)$  &  0.74 $\pm$ 0.03~\protect\cite{pdg96}&  $0.66$ & 0.68\\\hline
$\xi(1.085\pm 0.045) $& $0.88 \pm 0.18$~\protect\cite{argus93} & 
        0.93& 0.93$^\dagger$\\\hline
$\xi(1.18\pm 0.045) $& $0.93 \pm 0.12$~\protect\cite{argus93} & 
        0.86 & 0.86$^\dagger$\\\hline
$\xi(1.28\pm 0.050)$ & $0.68 \pm 0.06$~\protect\cite{argus93} & 
        0.80 & 0.79$^\dagger$\\\hline
$\xi(1.36\pm 0.050)$ & $0.66 \pm 0.06$~\protect\cite{argus93} & 
        0.75 & 0.75$^\dagger$\\\hline
$\xi(1.45\pm 0.045)$ & $0.58 \pm 0.08$~\protect\cite{argus93} & 
        0.71 & 0.70$^\dagger$\\\hline
$\rho^2 $    & $\displaystyle \begin{array}{l}
                                0.91 \pm 0.15 \pm 0.06 \\
                                1.53 \pm 0.36 \pm 0.14
                                \end{array}$~\cite{cesr96} 
                              & 0.89 & 0.91 \\\hline
$f_{B_s}\,$(GeV) & $0.195 \pm 0.035$~\cite{hqlat} & 0.220 & 0.220\\\hline
$f_{B_s}/f_B$ & $1.14 \pm 0.08 $~\cite{hqlat} & 1.071 & 1.071 \\\hline
$f_{D}\,$(GeV) & $0.200 \pm 0.030 $~\cite{hqlat} & 0.344 & 0.344 \\\hline
$f_{D_s}\,$(GeV) &  $ 0.220 \pm 0.030$~\cite{hqlat}  & 0.363 & 0.363 \\\hline
$f_{D_s}/f_D$ & $1.10\pm 0.06$~\cite{hqlat}& 1.054 & 1.054 \\\hline
\end{tabular}
\caption{A comparison of our calculated results with available data when we
require $f_B=0.205\,$GeV, which is the upper bound estimated in
Ref.~\protect\cite{hqlat}, and use Eq.~(\protect\ref{phia}).  (See
Table.~\protect\ref{tablea} for additional remarks and an explanation of the
symbols.)
\label{tablec}}
\end{table}

\end{document}